\documentclass[angeo]{copernicus}
   \def\ang{\AA}

   \def\gapprox{\lower.4ex\hbox{$\;\buildrel >\over{\scriptstyle\sim}\;$}}
   \def\lapprox{\lower.4ex\hbox{$\;\buildrel <\over{\scriptstyle\sim}\;$}}

\begin{document}

\title{4D Modeling of CME Expansion and EUV Dimming Observed with STEREO/EUVI}

\author[]{Markus J. Aschwanden}

\affil[]{Solar and Astrophysics Laboratory, Lockheed Martin ATC,
	3251 Hanover St., Bldg. 252, Org. ADBS, Palo Alto, CA 94304, USA}

\runningtitle{CME Expansion and EUV Dimming}

\runningauthor{Aschwanden}

\correspondence{Aschwanden\\ (email: aschwanden@lmsal.com)}

\received{}
\revised{}
\accepted{}
\published{}

\firstpage{1}

\maketitle

\begin{abstract}
This is the first attempt to model the kinematics of a CME launch
and the resulting EUV dimming quantitatively with a self-consistent model.
Our 4D-model assumes self-similar expansion of a spherical CME geometry
that consists of a CME front with density compression and a cavity with
density rarefaction, satisfying mass conservation of the total CME
and swept-up corona. The model contains 14 free parameters and is
fitted to the 2008 March 25 CME event observed with
STEREO/A and B. Our model is able to reproduce the observed CME expansion 
and related EUV dimming during the initial phase from 18:30 UT
to 19:00 UT. The CME kinematics can be characterized by a constant 
acceleration (i.e., a constant magnetic driving force). While the 
observations of EUVI/A are consistent with a spherical bubble geometry,
we detect significant asymmetries and density inhomogeneities with EUVI/B.  
This new forward-modeling method demonstrates how the observed EUV dimming 
can be used to model physical parameters of the CME source region, 
the CME geometry, and CME kinematics.
\end{abstract}

\section{Introduction}

It has become a generally agreed concept that the EUV dimming 
observed during the onset of a {\sl coronal mass ejection (CME)} manifests
the coronal mass loss of the CME, and thus we basically expect a
one-to-one correlation between the detections of CMEs and EUV dimmings,
unless there exist special circumstances. For instance, the CME could
originate behind the limb, in which case the EUV dimming is obscured,
or the CME could start in the upper corona, where there is little
EUV emission because of the gravitational stratification. The latter
case would imply very low masses compared with a CME that originates 
at the base of the corona, i.e., $\approx 10\%$ at two thermal scale heights. 
However, there exists a case with an average CME mass that did not
leave any footprints behind in EUV (Robbrecht et al. 2009).
A statistical study on the simultaneous detection of EUV dimmings and
CMEs has recently been performed by Bewsher et al. (2008).
This study based on SOHO/CDS and LASCO data confirms a 55\% association
rate of dimming events with CMEs, and vice versa a 84\% association rate
of CMEs with dimming events. Some of the non-associated events may be
subject to occultation, CME detection sensitivity, or incomplete temperature
coverage in EUV and soft X-rays.  Perhaps the CME-dimming association rate 
will reach 100\% once the STEREO spacecraft arrive at a separation of 
$180^\circ$ and cover all equatorial latitudes of the Sun.

A number of studies have been carried out by using the detection 
of coronal dimming to identify CME source regions, focusing on transient 
coronal holes caused by filament eruptions (Rust 1983; Watanabe et al.~1992), 
EUV dimming at CME onsets (Harrison 1997; Aschwanden et al.~1999), 
soft X-ray dimming after CMEs (Sterling \& Hudson 1997),
soft X-ray dimming after a prominence eruption (Gopalswamy \& Hanaoka 1998),
simultaneous dimming in soft X-rays and EUV during CME launches
(Zarro et al.~1999; Harrison \& Lyons 2000; Harrison et al.~2003),
determinations of CME masses from EUV dimming from spectroscopic data
(Harrison \& Lyons 2000; Harrison et al.~2003) or from EUV imaging data
(Zhukov and Auchere 2004; Aschwanden et al.~2009b).
All these studies support the conclusion that dimmings in the corona (either
detected in EUV, soft X-rays, or both) are unmistakable signatures of CME
launches, and thus can be used vice versa to identify the mutual phenomena.

In this study here we attempt for the first time to model the kinematics
of a CME and the resulting EUV dimming quantitatively, which provides
us unique physical parameters of the CME source region and on the 
CME kinematics in the initial acceleration phase.

\section{Model Assumptions}

The dynamics of a CME can often be
characterized by a rapid expansion of a magnetically unstable coronal
volume that expands from the lower corona upward into the heliosphere.
Different shapes have been used to approximately describe the 3D geometry
of a CME, such as a spherical bubble, an ice-cone, a crescent,
or a helical flux rope, which expand in a self-similar fashion and 
approximately maintain the aspect ratio in vertical and horizontal 
directions during the initial phase of the expansion. Here we develop
a four-dimensional (4D=3D+T) model that describes the 3D evolution of 
the CME geometry in time (T) in terms of 4D electron density distributions 
$n_e(x,y,z,t)$ that allow us also to predict and forward-fit a 
corresponding EUV intensity image data cube $I_{EUV}(x,y,t)$ in an 
observed wavelength.

For the sake of simplicity we start in our model here with the simplest case, 
assuming: (1) spherical 3D geometry for the CME front and cavity; 
(2) self-similar expansion in time; (3) density compression in the CME front 
and adiabatic volume expansion in the CME cavity; (4) mass conservation for 
the sum of the CME front, cavity, and external coronal volume; 
(5) hydrostatic (gravitational stratification) or super-hydrostatic 
density scale heights; (6) line-tying condition for the magnetic field
at the CME base; and (7) a magnetic driving force that is constant 
during the time interval of the initial expansion phase. 
This scenario is consistent with the traditional characterization of a typical
CME morphology in three parts, including a CME front (leading edge), a cavity,
and a filament (although we do not model the filament part). The expanding 
CME bubble sweeps up the coronal plasma that appears as a bright rim at
the observed ``CME front'' or leading edge. The interior of the CME bubble 
exhibits a rapid decrease in electron density due to the adiabatic expansion, 
which renders the inside of the CME bubble darker in EUV and appears as 
the observed ``CME cavity''.
The assumption of adiabatic expansion implies no mass and energy exchange
across the outer CME boundary, and thus is consistent with the assumption
of a low plasma $\beta$-parameter in the corona with perfect magnetic 
confinement, while the CME bubble will become leaking in the outer 
corona and heliosphere, where the plasma $\beta$-parameter exceeds unity
(not included in our model here). 

\section{Analytical Model}

A spherical 3D geometry can be characterized by one single free parameter,
the radius $R$ of the sphere. The self-similar expansion maintains the
spherical shape, so the boundary of the CME bubble can still be parameterized
by a single time-dependent radius $R(t)$. The time-dependence of the CME
expansion is controlled by magnetic forces, e.g., by a Lorentz force or
hoop force. For sake of simplicity we assume a constant force during the
initial phase of the CME expansion, which corresponds to a constant
acceleration $a_R$ and requires three free parameters ($R_0, v_R, a_R$) to 
characterize the radial CME expansion, 
\begin{equation}
	R(t) = R_0 + v_R (t - t_0) 
                   + {1 \over 2} a_R (t - t_0)^2 \quad {\rm for}\ t>t_0 \ ,
\end{equation}
where $R_0=R(t=t_0)$ is the initial radius at starting time $t_0$,
$v_R$ is the initial velocity and
$a_R$ is the acceleration of the radial expansion.

For the motion of the CME centroid at height $h(t)$ we assume a similar 
quadratic parameterization,
\begin{equation}
	h(t) = h_0 + v_h (t - t_0)
                   + {1 \over 2} a_h (t - t_0)^2 \quad {\rm for}\ t>t_0 \ ,
\end{equation}
where $h_0=h(t=t_0)$ is the initial height at starting time $t_0$,
$v_h$ is the initial velocity and
$a_h$ is the acceleration of the vertical motion. 
This parameterization is consistent with a fit to a theoretical MHD
simulation of a breakout CME (Lynch et al.~2004) as well as with 
kinematic fits to observed CMEs (Byrne et al.~2009).

\begin{figure}[t]
\vspace*{2mm}
\begin{center}
\includegraphics[width=8.3cm]{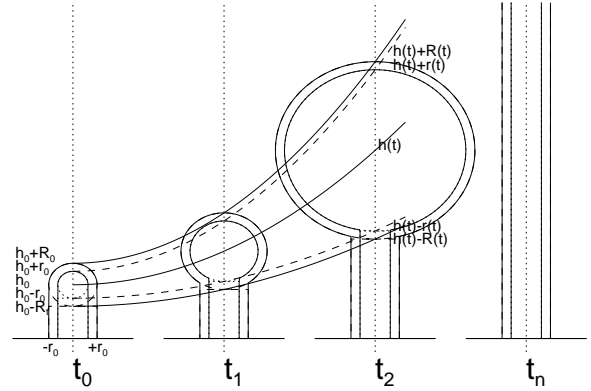}
\end{center}
\caption{The definition of the self-similar geometry of the CME model
is depicted for four different times, consisting of a cyclidrical base
and a spherical shell. The height of the centroid of the spherical CME 
volume is $h(t)$, the outer radius of the CME sphere is $R(t)$, and 
the inner radius of the CME front is $r(t)$. These parameters increase 
quadratically with time. The circular footpoint area of the CME with 
radius $r_0$ stays invariant during the self-similar expansion in order 
to satisfy the line-tying condition of the coronal magnetic field at the 
footpoints.}
\end{figure}

Further we constrain the CME geometry with a cylindrical footpoint
area of radius $r_0$, which connects from the solar surface to the
lowest part of the spherical CME bubble. In order to ensure magnetic
line-tying at the footpoints, the CME bubble should always be located
above the cyclidrical footpoint base, which requires that the initial height 
satisfies $h_0 > r_0$ and the acceleration constants are $a_h > a_r$.
We visualize the model geometry in Fig. 1. 

The time-invariant CME footprint area allows us a simple mass estimate 
of the CME from the cylindrical volume integrated over a vertical 
scale height, since the spherical CME bubble will eventually move 
to large heights with no additional mass gain (at time $t_n \gg t_0$,
see right-hand panel in Fig.1). 

Assuming adiabatic expansion inside the CME cavity, the electron density
in the confined plasma decreases reciprocally to the expanding volume, i.e.,
\begin{equation}
	n_e(t)=n_0 \left( {r(t) \over r_0} \right)^{-3} \ ,
\end{equation}
so it drops with the third power as a function of time from the initial
value $n_0$ (of the average density inside the CME). 

For the mass distribution inside the CME we distinguish between a
compression region at the outer envelope, containing the CME front,
and a rarefaction region in the inside, which is also called CME cavity.
We define an average width $w_0$ of the CME front that is assumed
to be approximately constant during the self-similar expansion of the CME.
While the radius $R(t)$ demarcates the outer radius of the CME front,
we denote the inner radius of the CME front or the radius of the cavity
with $r(t)$, 
\begin{equation}
	 r(t)=R(t)-w_0 	 	\ ,
\end{equation}
which has an initial value of $r_0 =r(t=t_0) =R_0-w_0$. 
The total volume $V_0$ of the CME is composed of a spherical volume 
with radius $R(t)$ and the cylindrical volume beneath the CME with a
vertical height of $(h_0-r_0)$,
\begin{equation}
	 V(t) = {4 \over 3} \pi R(t)^3 + \pi R_0^2 [h(t)-R(t)] \ ,
\end{equation}
which has an initial volume value of $V_0$,
\begin{equation}
	V_0 = V(t=t_0) = {4 \over 3} \pi R_0^3 + \pi R_0^2 [h_0 - R_0] \ .
\end{equation}
The volume of the CME front $V_F(t)$ is essentially the difference between
the outer and inner sphere (neglecting the cylindrical base at the footpoint)
\begin{equation}
	V_F(t) \approx {4 \over 3} \pi R(t)^3 - {4 \over 3} \pi r(t)^3 \ ,
\end{equation}
while the volume $V_C$ of the cavity is,
\begin{equation}
	V_C(t) \approx {4 \over 3} \pi r(t)^3  .
\end{equation}

We have now to define the time-dependent densities in the CME, for both
the CME front, which sweeps up plasma during its expansion, as well as
for the CME cavity, which rarifies due to the adiabatic expansion.
The total mass $m_E(t)$ of the plasma that is swept up from the external 
corona in a CME corresponds to the total CME volume $V(t)$ minus the
initial volume of the CME cavity,
\begin{equation}
	\begin{array}{ll}
	m_E(t)  &=m_p <n_E> [V(t)-V_C(t=t_0)] \\
		&=m_p <n_E> {4 \over 3} \pi [R(t)^3 - r_0^3] 
	\end{array} \ ,
\end{equation}
where $m_p$ is the mass of the hydrogen atom and $<n_E>$ is the 
electron density in the external corona averaged over the CME volume.
The same mass has to be contained inside the volume $V_F$ of the CME front,
\begin{equation}
	\begin{array}{ll}
	m_E(t)	&=m_p <n_F> V_F(t) 			\\
		&= m_p <n_F> {4 \over 3} \pi [R(t)^3 - r(t)^3]
	\end{array} \ ,
\end{equation}
Thus, mass conservation yields a ratio of the average electron density
$<n_F>$ in the CME front and the average external density $<n_E>$ of 
\begin{equation}
	q_{n,front}(t) = {<n_F> \over <n_E>} 
	       = {R(t)^3 - r_0^3 \over R(t)^3 - r(t)^3} \ . 
\end{equation}
This density ratio amounts to unity at the initial time, i.e., $q_n(t=t_0)=1$
and monotonically increases with time. The maximum value of the density jump 
in MHD shocks derived from the {\sl Rankine-Hugoniot relations} 
(e.g., Priest 1982) is theoretically $q_{n,max}=4$. 
Fast CMEs are expected to be supersonic and will have a higher compression 
factor at the CME front than slower CMEs. Thus we keep the maximum 
compression factor $q_{n,max}$ as a free parameter, keeping in mind that
physically meaningful solutions should be in the range of 
$1 \lapprox q_{n,max} \lapprox 4$.
Since we prescribe both the width $w_0$ of the CME front as well
as a maximum density compression ratio $q_{n,max}$ we obtain a definition
of the critical value $\rho(t)$ for the cavity radius $r(t)$ when the
prescribed maximum density compression $q_{n,max}$ is reached (using
Eq. 11), 
\begin{equation}
	q_{n,max} = {R(t)^3 - r_0^3 \over R(t)^3 - \rho(t)^3} \ ,
\end{equation}
which yields the critical radius $\rho(t)$,
\begin{equation}
	\rho(t) = \left[ R(t)^3 - {(R(t)^3-r_0^3) \over q_{n,max}} 
	\right]^{1/3} \ .
\end{equation}
Therefore, only plasma outside this critical radius $\rho(t)$ can be
compressed in the CME front, while the plasma inside this critical
radius dilutes by adiabatic expansion and forms the cavity, yielding 
an average density ratio $q_{n,cav}$ inside the cavity (according to Eq. 3),
\begin{equation}
	q_{n,cav}(t) = {n_{e,cav}(t) \over n_0} =
	\left\{
	\begin{array}{ll}
	{\left[ r_0 / \rho(t) \right]}^3 & {\rm for}\ \rho(t) \ge r(t) \\
	{\left[ r_0 /    r(t) \right]}^3 & {\rm for}\ \rho(t) <   r(t)
	\end{array}
	\right.
	\ .
\end{equation}

\begin{figure*}[t]
\vspace*{2mm}
\begin{center}
\includegraphics[width=16.6cm]{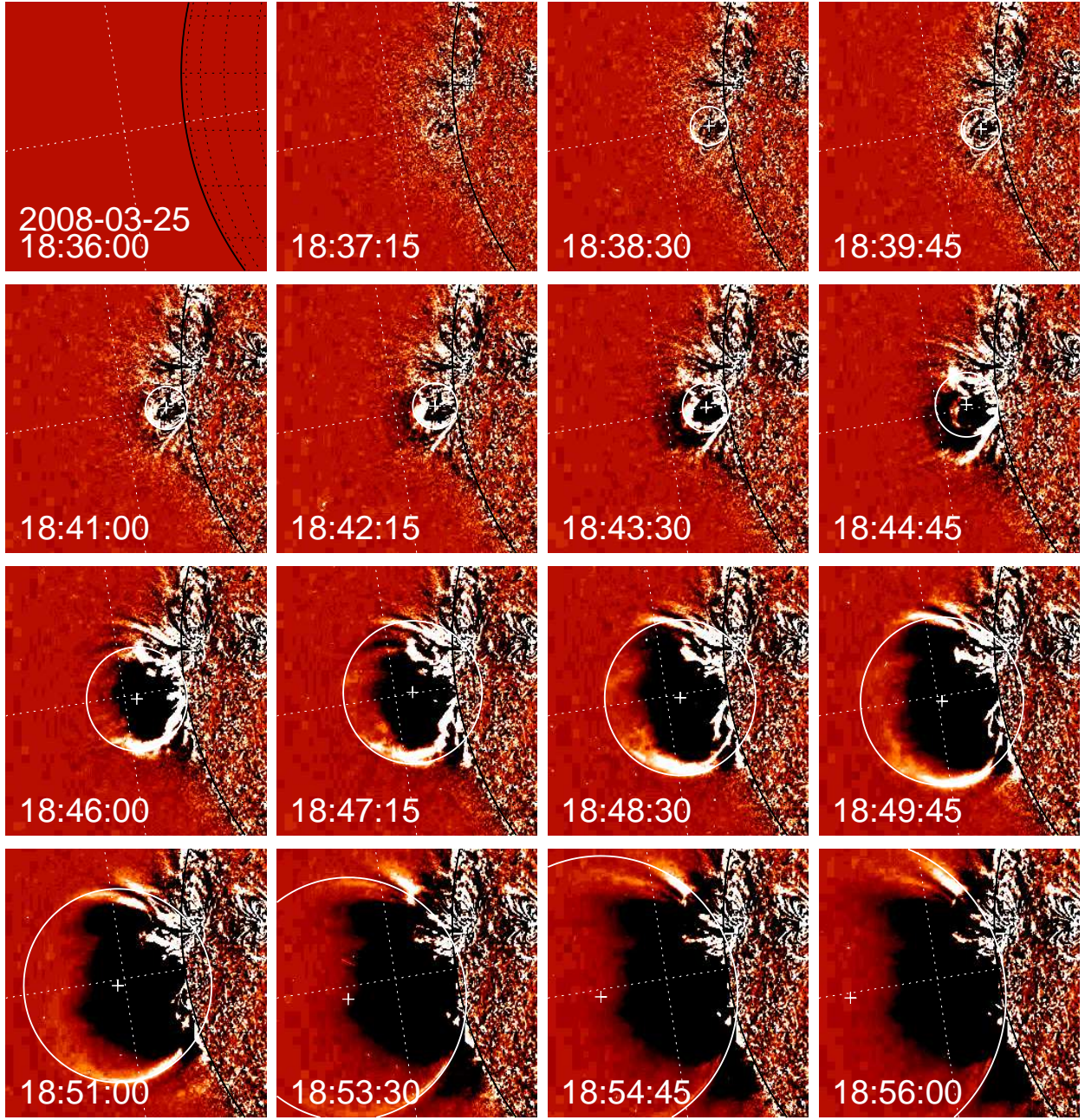}
\end{center}
\caption{Partial EUVI base-difference images from STEREO/A 171 \ang\ 
during the time interval of 2008 Mar 25, 18:36-18:56 UT (with the initial 
image at 18:36 UT subtracted). The envelope of the lateral CME front is
fitted with a circle. The dashed lines mark the locations where
cross-sectional brightness profiles are extracted for fitting,
shown in Fig. 5.}
\end{figure*}

\begin{figure*}[t]
\vspace*{2mm}
\begin{center}
\includegraphics[width=16.6cm]{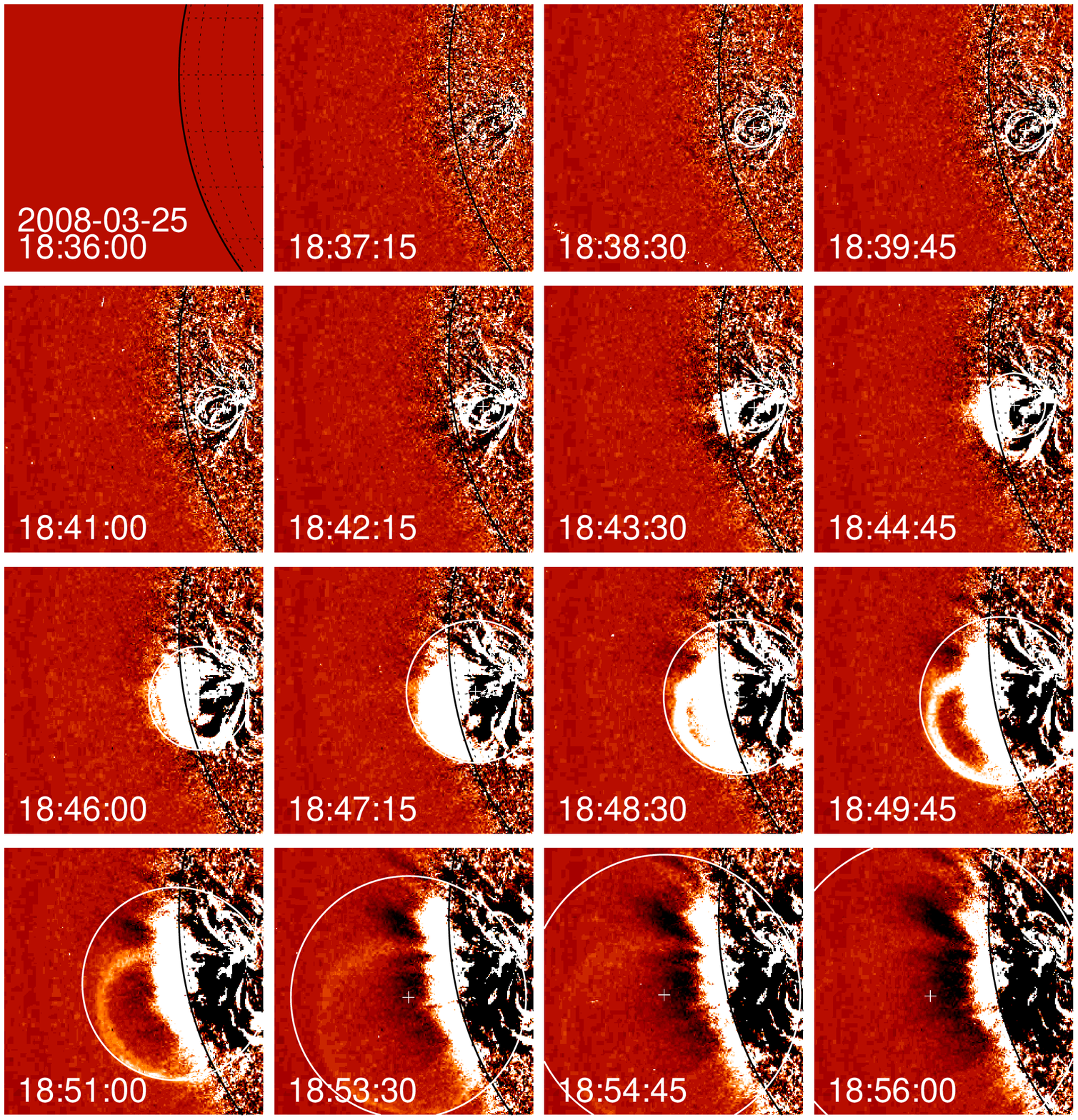}
\end{center}
\caption{EUVI base-difference images from STEREO/B 171 \ang\ 
in similar representation as for STEREO/A in Fig. 2.
Note that the CME origin is in the front of the solar disk in EUVI/B,
while it is behind the east limb for EUVI/A.}
\end{figure*}

Our numerical model of a spherical CME expansion has a total of 14 free
parameters: 3 positional parameters (the heliographic coordinates ($l,b$)
and height $h_0$ of the initial CME centroid position), 5 kinematic
parameters (starting time $t_0$, velocities $v_h, v_R$, accelerations 
$a_h, a_R$), 2 geometric parameters (initial radius $r_0$ and width $w_0$ 
of the CME front), and 4 physical parameters (coronal base 
density $n_0$, maximum density compression factor $q_{n,max}$ in the CME front, 
the mean coronal temperature $T_0$ at the observed wavelength filter),
and a vertical density scale height factor (or super-hydrostaticity factor)
$q_{\lambda}$ that expresses the ratio of the effective density scale height
to the hydrostatic scale height at temperature $T_0$).
The temperature $T_0$ defines the hydrostatic scale height $\lambda_T$ 
of the corona in the observed temperature range, which enters the definition
of the effective density scale height $\lambda$ (e.g., Eq. [3.1.16] in
Aschwanden 2004), 
\begin{equation}	
	\lambda = \lambda_T \ q_\lambda = 
	47 \left({T_0 \over 1 \ {\rm MK}}\right) q_\lambda \quad {\rm [Mm]}\ .
\end{equation}
Thus, assuming an exponentially stratified atmosphere (Eq.~15), 
a density compression factor $q_n(t) \le q_{n,max}$ in the CME front (Eq.~12), 
and adiabatic expansion inside the CME cavity (Eq.~14), 
we have the following time-dependent 3D density model:
\begin{equation}
	\begin{array}{ll}
	n_e(h,t) = 
	n_0 \exp{\left( - {h / \lambda} \right)} \times & \\
	  \qquad \qquad \times \left\{
	  \begin{array}{lll}
 	  1		&{\rm for}\ d    > R(t)     \\
 	  q_{n,front}(t)&{\rm for}\ r(t) < d < R(t) \\
 	  q_{n,cav}(t)  &{\rm for}\ d 	 < r(t) 
          \end{array}
	  \right. \ ,
	\end{array} 
\end{equation}
where $d$ is the distance of an arbitrary location with 3D coordinates 
$(x,y,z)$ to the instantaneous center position 
$[x_0(t), y_0(t), z_0(t)]$ of the CME,
\begin{equation}
	d = \sqrt{[x-x_0(t)]^2+[y-y_0(t)]^2+[z-z_0(t)]^2} \ ,
\end{equation}
which is located at height $h(t)$ vertically above the heliographic 
position ($l,b$).

\section{Forward-Fitting of Model to Observations}

\subsection{STEREO/EUVI Observations}

One CME event observed with STEREO that appears as a spherically
expanding shell, and thus is most suitable for fitting with our analytical
model, is the 2008-Mar-25, 18:30 UT, event. This CME occurred near the
East limb for spacecraft STEREO/Ahead, and was observed on the frontside
of the solar disk with spacecraft STEREO/Behind. Some preliminary analysis
of this event regarding CME expansion and EUV dimming can be found in 
Aschwanden et al. (2009a), the CME mass was determined in white light
with STEREO/COR-2 (Colaninno and Vourlidas 2009) and with STEREO/EUVI
(Aschwanden et al. 2009b), and the 3D geometry was modeled with 
forward-fitting of various geometric models to the white-light
observations (Thernisien et al. 2009; 
Temmer et al. 2009; Maloney et al. 2009; Mierla et al. 2009a,b). 
While most previous studies model the white-light emission of this CME,
typically a few solar radii away from the Sun, our model applies
directly to the CME source region in the lower corona, as observed
in EUV. We follow the method outlined in Aschwanden et al. (2009a). 
Our model also quantifies the geometry and kinematics of the CME,
as well as the EUV dimming associated with the launch of the CME.

\subsection{Fitting of Positional Parameters}

Figures 2 and 3 show sequences of 16 (partial) EUV images, 
simultaneously observed with STEREO/A and B with a cadence of 75 s during 
the time interval of 18:36-18:56 UT on 2008-Mar-25. In order to determine
the positional parameters of the CME as a function of time we 
trace the outer envelope of the CME bubble (by visual clicking of 
3 points) in each image and each spacecraft and fit a circle through
the 3 points in each image. 
The selected points for fitting the position of the CME bubble were
generally chosen in the brightest features of the lateral CME flanks,
but could not always been traced unambiguously in cases with
multiple flank features. In those cases we traced the features that
were closest to a continuously expanding solution.
The radii and y-positions of the circular fits are fully constrained
from the STEREO/A images, so that only the x-positions of the centroid
of the spherical shell need to be fitted in the epipolar STEREO/B images.
We note that the fits of the CME
bubble roughly agree with the envelopes of the difference flux
in the STEREO/B images initially (up to 18:48 UT), while there is
a discrepancy later on. Apparently the CME has a more complex
geometry than our spherical bubble model, which needs to be
investigated further.

This procedure yields the CME centroid 
positions $[x_A(t_i), y_A(t_i)]$ and $[x_B(t_i), y_B(t_i)]$ for
the time sequence $t_i, i=1,...,16$. The images in Fig. 2 and 3 
are displayed as a baseline difference (by subtracting a pre-CME image 
at 18:36 UT) to enhance the contrast of the CME edge. The circular fits 
to the CME outer boundaries are overlayed in Fig. 2 and 3.
Both images have been coaligned and rotated into epipolar coordinates
(Inhester et al. 2006), so that the y-coordinates of a corresponding
feature are identical in the spacecraft A and B images, while the x-coordinates
differ according to the spacecraft separation angle $\alpha_{sep}$, which
amounts to $\alpha_{sep}=47.17^\circ$ at the time of the CME event.
The epipolar coordinates measured from both spacecraft are then related 
to the heliographic longitude $l$, latitude $b$, and distance $r_c$ from Sun
center as follows,
\begin{equation}
	\begin{array}{l}
	x_A = r_c \cos(b) \sin(l_A) 		\\
	y_A = y_B = r_c \sin(b)		\\
	x_B = r_c \cos(b) \sin(l_A+\alpha_{sep}) 
	\end{array}	\ ,
\end{equation}
which can directly be solved to obtain the spherical (epipolar) coordinates 
$(l_A,b,r_c )$,
\begin{equation}
	\begin{array}{l}
	l_A = \arctan{ { \left( \sin{(\alpha_{sep})} / 
		(x_A/x_B - \cos{(\alpha_{sep})} \right) }}\\
	r_c = \sqrt{ y_A^2 + [x_A/\sin{(l_A)}]^2 }	\\
	b   = \arcsin{( y_A / r_c )}
	\end{array}
\end{equation}
Therefore, using stereoscopic triangulation, we can directly determine
the spherical coordinates $(l_i, b_i, r_{c,i})$, $i=1,...,16$ for all 16
time frames, as well as obtain the CME curvature radii $R(t_i)$  
from the circular fits to the CMEs. We plot the so obtained observables
$l_A(t)$, $b(t)$, $R(t)$, and $h(t)=r_c(t)-R_{\odot}$ in Fig. 3 and
determine our model parameters $l$ and $b$ from the averages.
We obtain a longitude of $l_A=-102.4^\circ \pm 0.9^\circ$ (for spacecraft 
STEREO/A), $l_B=l_A + \alpha_{sep} = -54.9^\circ \pm 0.9^\circ$ 
(for spacecraft STEREO/B), and a latitude $b=-8.8^\circ\pm0.6^\circ$. 
Thus, the CME source region is clearly occulted for STEREO/A.
These epipolar coordinates 
can be rotated into an ecliptic coordinate system by the tilt angle
$\beta_{AB}=2.66^{\circ}$ of the spacecraft A/B plane. Viewed from Earth, 
the longitude is approximately $l_E \approx -102.4 + \alpha_{sep}/2 
\approx -78.8^\circ$. Thus, the CME source region is $12^\circ$ behind
the East limb when seen from Earth. This explains why the EUV dimming
is seen uncontaminted from post-flare loops, which are seen by
STEREO/B but hidden for STEREO/A. 

\begin{figure}[t]
\vspace*{2mm}
\begin{center}
\includegraphics[width=8.3cm]{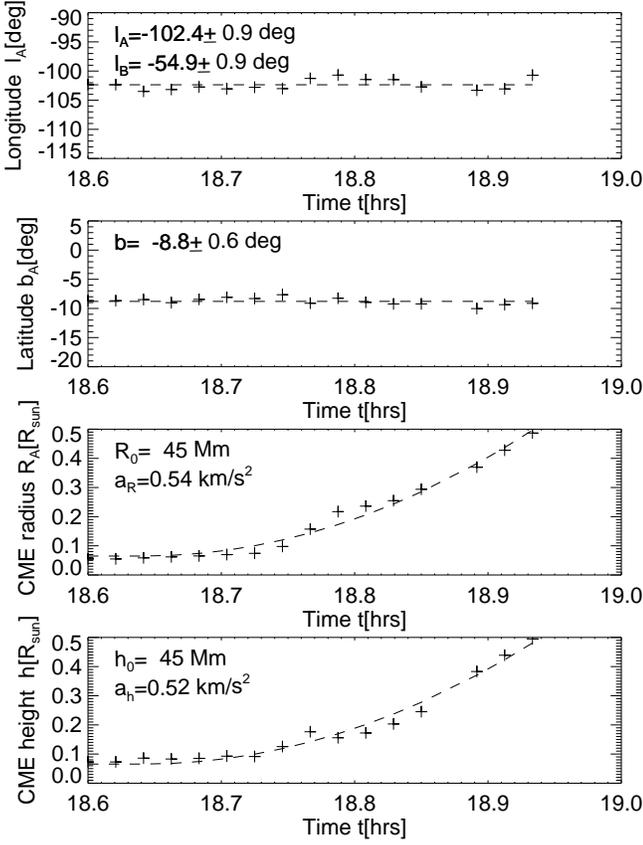}
\end{center}
\caption{Measurements of the heliographic longitude $l_A(t)$ (top panel) 
and latitude $b(t)$ (second panel) of the CME centroid, the radius $R(t)$
of the CME sphere (third panel), and CME centroid height $h(t)$ (bottom).
The average values are $l_A=-102.4^\circ\pm0.9^\circ$ and $b=-8.8^\circ
\pm 0.9^\circ$ for the heliographic position. The CME radius $R(t)$ and
height $h(t)$ are fitted with quadratic functions (dashed curves), 
yielding the constants
$t_0=18.64$ hrs (18:38 UT), $R_0=45$ Mm, $h_0=45$ Mm, and accelerations
$a_R=0.54$ km s$^{-2}$ and $a_h=0.52$ km s$^{-2}$ (see Eqs. 1-2).}
\end{figure}

\begin{figure*}[t]
\vspace*{2mm}
\begin{center}
\includegraphics[width=16.6cm]{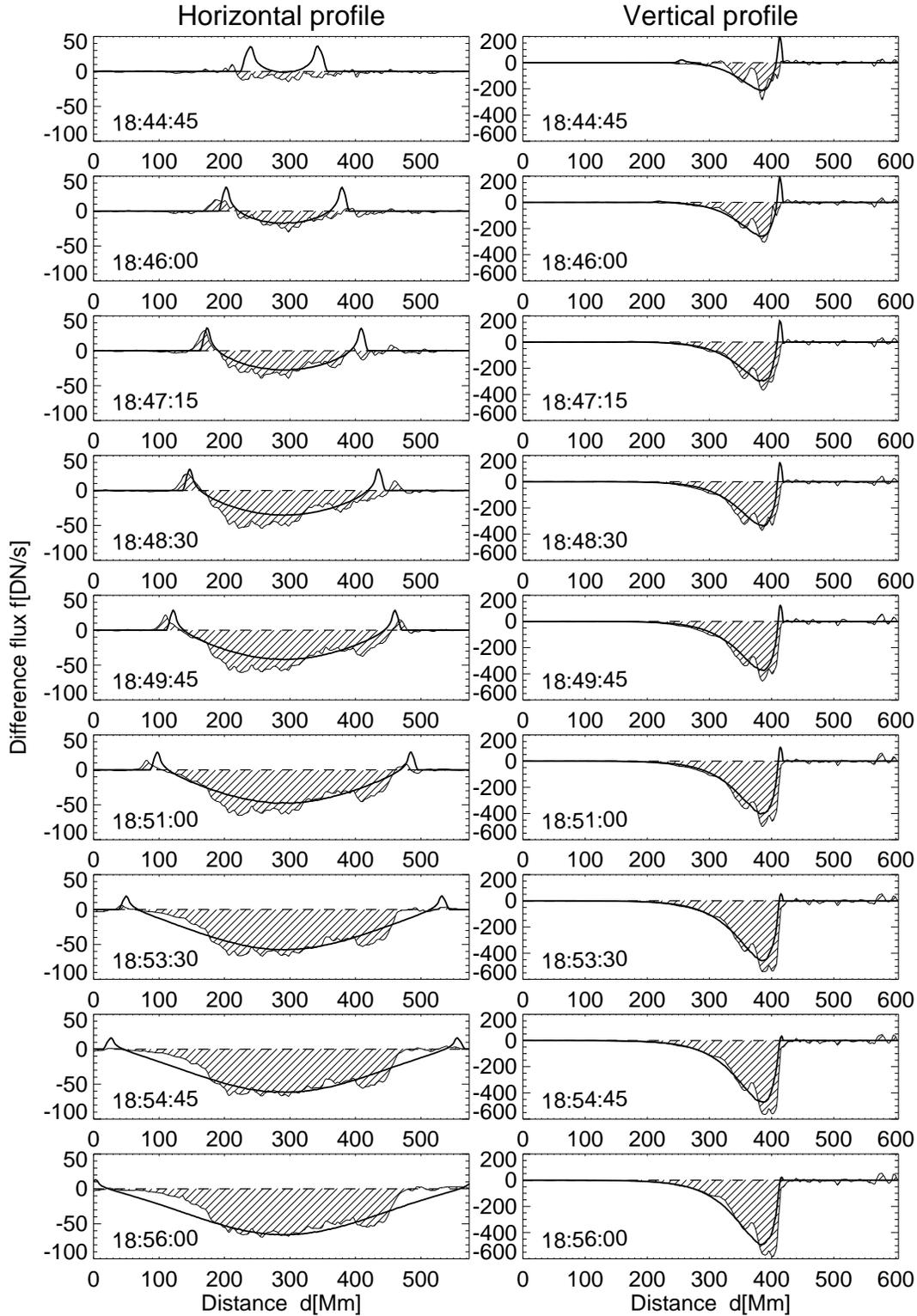}
\end{center}
\caption{Cross-sectional (baseline-subtracted) EUV brightness profiles
in horizontal direction above the limb (left-hand panels) and in vertical 
direction through the centroid of the dimming region (right-hand panels),
as marked with dashed lines in Fig.~2, for the 9 images during 
18:44:56$-$18:56:00 UT, for EUVI/A, 171 \ang . The observed dimming is
represented with hatched areas, while the best-fit model profiles are
plotted with thick linestyle.} 
\end{figure*}

\begin{figure*}[t]
\vspace*{2mm}
\begin{center}
\includegraphics[width=16.6cm]{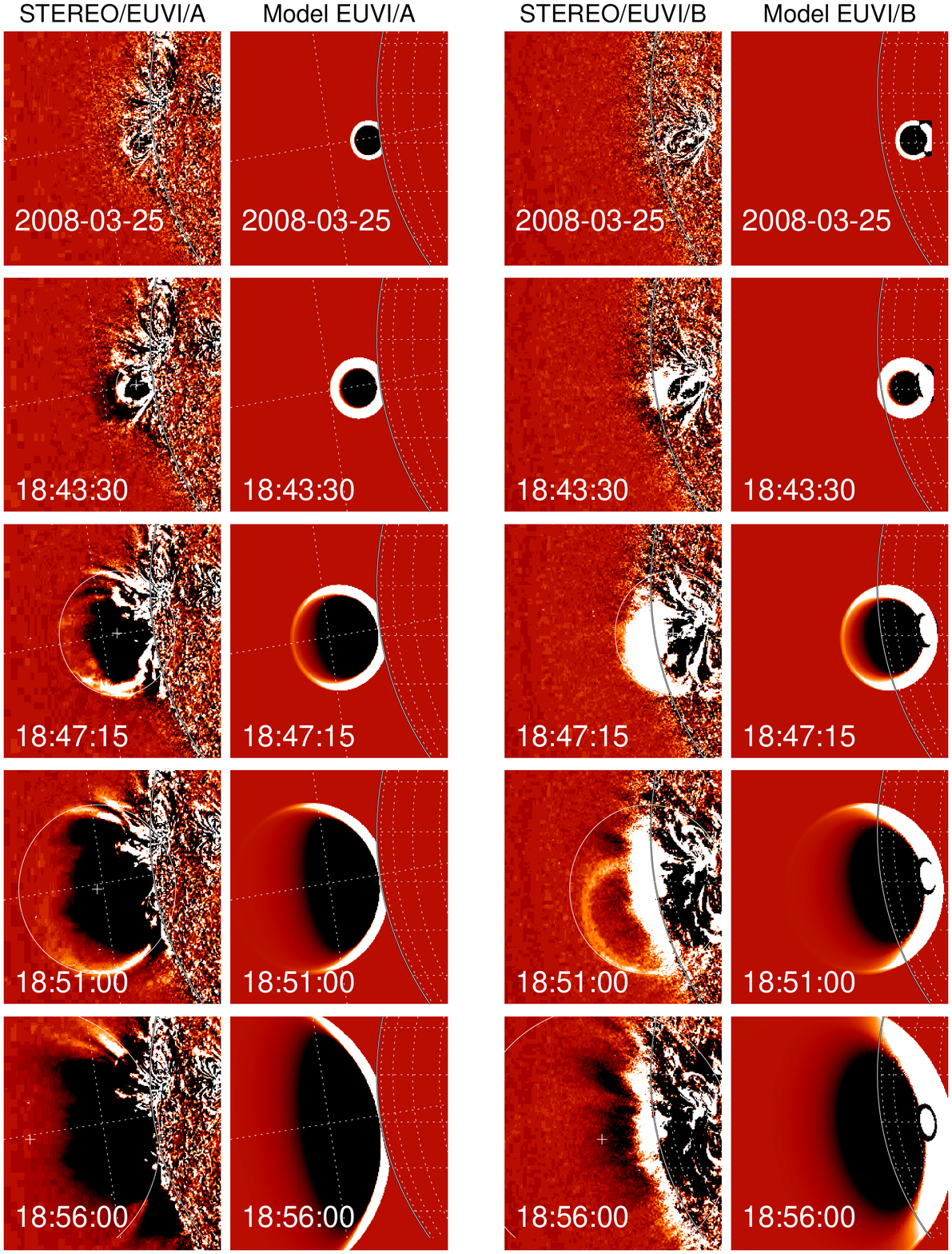}
\end{center}
\caption{Comparison of observed and simulated EUVI base-difference images 
at 5 times for the observations of STEREO/A 171 \ang\ (left two columns) 
and STEREO/B 171 \ang (right two columns). The pre-CME image at 18:36 UT
was subtracted in these base-difference images.}
\end{figure*}

\subsection{Fitting of Kinematic Parameters}

We plot also the observables $h(t)$ and $R(t)$  in Fig. 4 and
determine the model parameters $h_0$, $R_0$, $a_h$, $a_R$ by fitting 
the quadratic functions $R(t)$ (Eq.~1) and $h(t)$ (Eq.~2), for which
we obtain the starting time $t_0=18.64$ hrs (18:38 UT), the initial
CME radius $R_0=45$ Mm, the initial height $h_0=45$ Mm, and the
accelerations $a_R=0.54$ km s$^{-2}$ for the CME radius expansion,
and $a_h=0.52$ km s$^{-2}$ for the height motion. 
The initial velocity is found to be negligible ($v_R\approx 0$
and $v_h \approx 0$). We estimate the
accuracy of the acceleration values to be of order $\approx 10\%$,
based on the uncertainty of defining the leading edge of the CME.
Thus, we determined 9 out of the 14 free parameters of our model sofar. 

Note that the acceleration measured here refers to the very origin
of the CME in low coronal heights of $\lapprox 0.6$ solar radii 
observed in EUVI data.  The acceleration is expected to be initially high
and to rapidly decline further out, when the driving magnetic 
forces decrease at large altitudes. 
This explains why our values for the acceleration in low coronal
heights are significantly higher than measured further out in the
heliosphere, typically in the order of tens of m s$^{-1}$ in 
height ranges of 5-22 solar radii, as measured with SOHO/LASCO.
SOHO/LASCO reported even a slightly negative acceleration at
altitutes of 5-22 solar radii. The driving magnetic forces that 
accelerate a CME are clearly confined to much lower altitudes.

\subsection{Fitting of Geometric Parameters}

We model the 3D geometry of the CME bubble with the time-dependent
radius $R(t)$ and the width $w_0$ of the CME compression region. 
In Fig. 5 we show cross-sectional EUV brightness
profiles across the CME in horizontal direction (parallel to the
solar surface) and in vertical direction for the EUVI/A 171 \ang\ 
observations (indicated with dotted lines in Fig. 2). 
These baseline-subtracted profiles clearly show 
a progressive dimming with a propagating bright rim at the 
CME boundary, which corresponds to the density compression region 
at the lateral expansion fronts of the CME. The bright rims are
clearly visible in the images during 18:46$-$18:56 UT shown in Fig.~2.
The average width of the observed bright rims is $w_0 \approx 10$ Mm,
a value we adopt in our model.

\subsection{Fitting of Physical Parameters}

Finally we are left with the four physical parameters $T_0, q_{\lambda}, n_0$,
and $q_{n,max}$. Since we show here only data obtained with the 171 \ang\
filter, the mean temperature is constrained by the peak temperature
of the instrumental EUVI response function, which is at $T_0=0.96$ MK.
This constrains the thermal scale height to $\lambda_T = 47,000 \times
0.96 = 45,000$ km. 

The remaining free parameters $q_{\lambda}$, $n_0$, and $q_{n,max}$
need to be determined from best-fit solutions by forward-fitting of 
simulated EUV brightness images (or profiles, as shown in Fig.~5)
to observed EUV brightness images (or profiles).
The EUV emission measure in each pixel position $(x,y)$ can be
computed by line-of-sight integration along the $z$-axis in our
3D density cube $n_e(x,y,z)$ per pixel area $dA$ for each time $t_i$, 
\begin{equation}
	EM(x,y) = \int n_e^2(x,y,z) dz dA \ ,
\end{equation}
from which the intensity $I_{171}(x,y)$ in the model image in units
of DN s$^{-1}$ can be directly obtained by multiplying with the 
instrumental response function $R_{171}(T)$ of the 171 \ang\ filter,
\begin{equation}
	\begin{array}{ll}
	I_{171}(x,y)&=\int EM(x,y) R_{171}(T) dT 	\\
	            &\approx EM(x,y) R_{171}(T_0) \Delta T_{171} 
	\end{array}	\ ,
\end{equation}
where $R_{171}=3632 \times 10^{-44}$ DN s$^{-1}$ cm$^{3}$ MK$^{-1}$ 
and the FWHM of the 171 filter is $\Delta T_{171}=1.25-0.51=0.74$ MK.

In Fig. 6 we show best-fit solutions of horizontal and vertical
brightness profiles. The absolute flux level is proportional to
the coronal base density squared, which we obtain by minimizing
the mean flux difference between simulated and observed flux profiles.
We obtain a best-fit value of $n_0=6.5 \times 10^8$ cm$^{-3}$.
The super-hydrostaticity factor is most sensitive to the vertical
flux profile (Fig.~ 5, right-hand side panels), for which we find
a best-fit value of $q_{\lambda} = 1.45$. Thus, the average density
scale height in the CME region is slightly super-hydrostatic, as
expected for dynamic processes. These values are typical for quiet-Sun 
corona and active region conditions (see Figs. 6 and 10 in Aschwanden 
and Acton 2001). 

The last free parameter, the maximum density compression factor $q_{n,max}$,
affects mostly the brightness of the CME rims. Fitting the brightness
excess at the CME rims at those times where bright rims are visible
are consistent with a value of $q_{n,max} \approx 2$. 

\begin{table}[t]
\caption{Model parameters}
\vskip4mm
\centering
\begin{tabular}{ll}
\tophline
Parameter			& Best-fit value		\\
\middlehline
Spacecraft separation angle	& $\alpha_{sep}=47.17^\circ$	\\
Spacecraft plane angle to ecliptic & $\beta_{AB}=2.66^\circ$	\\
Heliographic longitude A	& $\l_A=-102.4^\circ\pm0.9^\circ$\\
Heliographic longitude B	& $\l_B=-54.9^\circ\pm0.9^\circ$\\
Heliographic longitude Earth	& $\l_E=-78.8^\circ\pm0.9^\circ$\\
Heliographic latitude  A,B 	& $\b  =-8.8^\circ\pm0.6^\circ$	\\
Start time of acceleration	& $t_0$=2008-Mar-25 18:38 UT	\\
Start time of modeling          & $t_1$=2008-Mar-25 18:36 UT	\\
End time of modeling            & $t_1$=2008-Mar-25 18:56 UT	\\
Initial height of CME center	& $R_0=45$ Mm			\\
Initial radius of CME 		& $h_0=45$ Mm			\\
Width of CME front 		& $w_0 \approx 10$ Mm		\\
Acceleration of vertical motion & $a_h=0.52$ m s$^{-2}$		\\
Acceleration of radial expansion& $a_R=0.54$ m s$^{-2}$		\\
Inital vertical velocity        & $v_h \approx 0$ m s$^{-1}$	\\
Inital expansion velocity       & $v_R \approx 0$ m s$^{-1}$	\\
Maximum density compression	& $q_{n,max} \approx 2.0$	\\
Corona/CME base density		& $n_0=6.5 \times 10^8$ cm$^{-3}$\\
Super-hydrostaticity factor	& $q_{\lambda} = 1.45$		\\
Mean Temperature (171 filter)	& $T_0=0.96$ MK			\\
Temperature width (171 filter)  & $\Delta T=\pm 0.37$ MK	\\
\bottomhline
\end{tabular}
\end{table}

\subsection{Comparison of Numerical Simulations with Observations}

After we constrained all 14 free parameters (listed in Table 1) 
of our analytical 4D model by fitting some observables, such as 
measured coordinates (Fig.~4) and cross-sectional horizontal 
and vertical brightness profiles (Fig.~5), we are now in the position
to inter-compare the numerically simulated images with the observed images,
as shown in Fig. 6 for 5 selected times, for both the STEREO/A and B
spacecraft. The comparison exhibits a good match for the extent of the
dimming region the the bright lateral rims, both extending over about
1.5 thermal scale heights above the solar surface. The base-difference
images of EUVI/A reveal a fairly symmetric CME (as the model is by design),
surrounded by spherical bright rims at the northern and southern
CME boundaries (as the model is able to reproduce it). 

The model, however, is less in agreement with the observed EUVI/B images.
The extent of the EUV dimming region matches relatively exactly,
although the observed dimming region is somewhat cluttered with bright
postflare loops that appear in the aftermath of the CME, which are
mostly hidden in the EUVI/A observations. The biggest discrepancy between
the model and the EUVI/B observations is the location of the brightest
rim of the CME boundary. The combination of projection effects and
gravitational stratification predict a brighter rim on the west side,
where we look through a longer and denser column depth tangentially 
to the CME bubble, which is not apparent in the observations of EUVI/B.
Instead, there is more bright emission just above the eastern limb that
cannot be reproduced by the model. Apparently there exists stronger density
compression on the eastern side of the CME bubble than the model predicts. 
Another inconsistency is the bright loop seen in EUVI/B at 18:51 UT, 
which does not match the surface of the modeled CME sphere as constrained 
by EUVI/A. Apparently, there are substantial deviations from a spherically 
symmetric CME bubble model that are visible in EUVI/B but not in EUVI/A. 
Perhaps a flux rope model could better fit the observations than a
spherical shell model. These discrepancies between the observations and 
our simple first-cut model provide specific constraints for a more complex 
model (with more free parameters) that includes inhomogeneities in the 
density distribution of the CME.

\subsection{Estimate of the CME Mass}

Our model allows us, in principle, to estimate the CME mass by integrating
the density $n_e(x,y,z,t)$ over the entire CME sphere, which is of course
growing with time, but expected to converge to a maximum value once the
CME expands far out into the heliosphere. A simple lower limit can 
analytically be obtained by integrating the density in the cylindrical
volume above the footpoint area, 
\begin{equation}
	M_{CME} = m_p \int n_e(x,y,z) dV_{CME} 
		\approx m_p \pi R_0^2 n_0 \lambda_T q_{\lambda} \ .
\end{equation}
From our best-fit values $R_0=45$ Mm, $q_{\lambda}=1.45$, $n0=6.5 \times
10^8$ cm$^{-3}$ and the thermal scale height of $\lambda_T=47$ Mm, 
we obtain a lower limit of $M_{CME} \ge 0.47 \times 10^{15}$ g.
However, this CME appears to expand in a cone-like fashion in the lowest
density scale height, so the total volume and mass is likely to be about
a factor of $\approx 2$ higher. Moreover, the mass detected in 171 \ang\
amounts only to about a third of the total CME mass (Aschwanden et al.
2009b), so a more realistic estimate of the total CME mass is about a
factor 6 higher than our lower limit, i.e., 
$M_{CME} \approx 3 \times 10^{15}$ g, which brings it into the ballpark
of previous CME mass determinations of this particular event, i.e., 
$m_{CME} = 2.9 \times 10^{15}$ g from STEREO/COR-2 white-light
observations (Colaninno and Vourlidas 2009), or
$m_{CME} = (4.3 \pm 1.4) \times 10^{15}$ g from STEREO/EUVI 
observations (Aschwanden et al. 2009b). 

\section{Conclusions}

We developed an analytical 4D model that simulates the CME expansion and 
EUV dimming in form of a time sequence of EUV images that can directly
be fitted to stereoscopic observations of STEREO/EUVI. The dynamic
evolution of the CME is characterized by a self-similar adiabatic
expansion of a spherical CME shell, containing a bright front with
density compression and a cavity with density rarefaction, satisfying
mass conservation of the CME and ambient corona. We forward-fitted this
model to STEREO/EUVI observations of the CME event of 2008-Mar-25 and
obtained the following results:

\begin{enumerate}
\item{The model is able to track the true 3D motion and expansion of the
CME by stereoscopic triangulation and yields an acceleration of 
$a \approx 0.54$ m s$^{-2}$ for both the vertical centroid motion and
radial expansion during the first half hour after CME launch.}

\item{Fitting the EUV dimming region of the model to the data mostly
constrains the coronal base density in the CME region 
($n_0 = 6.5 \times 10^8$ cm$^{-3}$) and the density scale height,
which was found to be super-hydrostatic by a factor of $q_{\lambda}=1.45$.}

\item{The average CME expansion speed during the first 10 minutes is
approximately $<v>\approx 400$ km s$^{-1}$, similar to the propagation
speeds measured for EIT waves in the initial phase (i.e.,
$v \approx 460$ km s$^{-1}$, Veronig et al. 2008;
$v \approx 475$ km s$^{-1}$, Long et al. 2008;
$v \approx 250$ km s$^{-1}$, Patsourakos et al. 2009),
and thus the CME expansion speed seems to be closely related to
the associated propagation kinematics of EIT waves.}

\item{The derived base density, scale height, and footpoint area constrain
the CME mass, but accurate estimates require a more complete temperature 
coverage with other EUV filters (e.g., see Aschwanden et al. 2009b).}

\item{The width and density compression factor of the CME front are
also constrained by our model, but accurate values require a perfectly
homogeneous CME shell.}

\item{While the spherical shell geometry reproduces the EUVI/A observations
well, significant deviations are noted in the EUVI/B observations, indicating
substantial inhomogeneities in the CME shell, possibly requiring a
hybrid model of bubble and flux rope geometries.}
\end{enumerate}

The most important conclusion of this modeling study is that EUV dimming
can be understood in a quantitative manner and that it provides a direct 
measurement of the coronal mass loss of a CME.
This exercise has shown that the spherical shell geometry can reproduce
a number of observed features of an evolving CME, which constrains the
physical and kinematic parameters of the initial phase of the CME launch,
but reveals also significant deviations that require a modification of
the idealized homogeneous spherical shell model.
The method of analytical 4D models with forward-fitting to stereoscopic
EUV images appears to be a promising tool to investigate quantitatively
the kinematics of CMEs. Combining with simultaneous magnetic or MHD modeling
may further constrain the physical parameters and ultimately provide the
capability to discriminate between different theoretical CME models. 
Full 4D modeling of the initial CME expansion may also provide a
self-consistent treatment of EIT waves and CME expansion (e.g., Chen et al. 
2005), for which we find similar kinematic parameters. 

\begin{acknowledgements}
This work is supported by the NASA STEREO under NRL contract N00173-02-C-2035.
The STEREO/ SECCHI data used here are produced by an international consortium of
the Naval Research Laboratory (USA), Lockheed Martin Solar and Astrophysics Lab
(USA), NASA Goddard Space Flight Center (USA), Rutherford Appleton Laboratory
(UK), University of Birmingham (UK), Max-Planck-Institut f\"ur
Sonnensystemforschung (Germany), Centre Spatiale de Li\`ege (Belgium), Institut
d'Optique Th\'eorique et Applique (France), Institute d'Astrophysique Spatiale
(France).
The USA institutions were funded by NASA; the UK institutions by
the Science \& Technology Facility Council (which used to be the Particle
Physics and Astronomy Research Council, PPARC); the German institutions by
Deutsches Zentrum f\"ur Luft- und Raumfahrt e.V. (DLR); the Belgian institutions
by Belgian Science Policy Office; the French institutions by Centre National
d'Etudes Spatiales (CNES), and the Centre National de la Recherche Scientifique
(CNRS). The NRL effort was also supported by the USAF Space Test Program and
the Office of Naval Research.
\end{acknowledgements}

\section*{References} 

\def\ref#1{\par\noindent\hangindent1cm {#1}}
\def\aaa{{\it Astron. Astrophys.}\ }
\def\aass{{\it Astron. Astrophys. Suppl. Ser.}\ }
\def\apj{{\it Astrophys. J.}\ }
\def\grl{{\it Geophys. Res. Lett.}\ }
\def\jgr{{\it J. Geophys. Res.}\ }
\def\ssr{{\it Space Scien. Rev.}\ }
\def\sp{{\it Solar Phys.}\ }
\small

\ref Aschwanden, M.J. 2004, {\sl Physics of the Solar Corona.
	An Introduction}, Praxis: Chichester/UK and Springer:Berlin. 
\ref Aschwanden, M.J., Fletcher, L., Schrijver, C., \& Alexander, D.
        1999, ApJ 520, 880.
\ref Aschwanden, M.J. and Acton, L.W. 2001, ApJ 550, 475. 
\ref Aschwanden, M.J., Nitta, N.V., Wuelser, J.P., and Lemen 2009a, 
	\sp 256, 3. 
\ref Aschwanden, M.J., Nitta, N.V., Wuelser, J.P., Lemen, J.R., Sandman, A.,
	Vourlidas, A., and Colaninno, R. 2009b, ApJ (subm).
\ref Bewsher, D., Harrison, R.A., \& Brown, D.S. 2008, A\&A 478, 897.
\ref Byrne, J.P., Gallagher, P.T., McAteer, R.T.J., and Young, C.A.
	2009, A\&A 495, 325. 
\ref Colaninno, R.C. and Vourlidas, A. 2009, ApJ (in press)
\ref Gopalswamy, N. \& Hanaoka, Y. 1998, ApJ 498, L179.
\ref Harrison, R.A. 1997, Proc. 31st ESLAB Symp., Correlated Phenomena
        at the Sun, in the Heliosphere and in Geospace, ESA SP-415, 121.
\ref Harrison, R.A. \& Lyons, M. 2000, A\&A 358, 1097.
\ref Harrison, R.A., Bryans, P., Simnett, G.M., \& Lyons, M. 2003,
        A\&A 400, 1071.
\ref Inhester, B. 2006, {\sl Stereoscopy basics for the STEREO mission},
	eprint arXiv: astro-ph/0612649 .
\ref Long, D.M., Gallagher, P.T., McAteer, R.T.J., and Bloomfield, D.S.
	2008, ApJ 680, L81. 
\ref Lynch, B.J., Antiochos, S.K., MacNeice, P.J., Zurbuchen, T.H.,
	and Fisk, L.A. 2004, ApJ 617, 589. 
\ref Maloney, S.A., Gallagher, P.T., and McAteer, R.T.J. 2009, SP 256, 149.
\ref Mierla, M., Inhester, B., Marque,C., Rodriguez, L., Gissot, S.,
	Zhukov, A.N., Berghmans, D., and Davila J. 2009a, SP (subm).
\ref Mierla, M. et al. 2009b, Ann.Geophys. (in preparation).
\ref Patsourakos, S. and Vourlidas, A. 2009, ApJ 700, L182.
\ref Priest, E.R. 1982, {\sl Solar Magnetohydrodynamics}, Geophysics
	and Astrophysics Monographs, Vol. 21, D. Reidel Publishing Company,
	Dordrecht, Netherlands.
\ref Robbrecht, E., Patsourakos, S., and Vourlidas, A. 2009, ApJ (in press). 
\ref Rust, D.M. 1983, SSRv 34, 21.
\ref Sterling, A.C. \& Hudson, H.S. 1997, ApJ 491, L55.
\ref Temmer, M., Preiss, S., and Veronig, A.M. 2009, SP 256, 183.
\ref Thernisien, A., Vourlidas, A., and Howard, R.A. 2009, SP (in press).
\ref Veronig, A.M., Temmer, M., and Vrsnak, B. 2008, ApJ 681, L113. 
\ref Watanabe, T., Kozuka,Y., Ohyama, M., Kojima, M., Yamaguchi, K.,
        Watari, S.I., Tsuneta, S., Jeselyn, J.A., Harvey, K.L.,
        Acton, L.W., \& Klimchuk J.A. 1992, PASJ 44, L199.
\ref Zarro, D.M., Sterling, A.C., Thompson, B.J., Hudson, H.S., \& Nitta, N.
        1999, ApJ 520, L139.
\ref Zhukov, A.N. and Auchere, F. 2004, \aaa 427, 705.



\end{document}